\documentclass[aps,prb,twocolumn,showpacs,showkeys]{revtex4}%
\usepackage{amsfonts}
\usepackage{amsmath}
\usepackage{amssymb}
\usepackage{graphicx}%
\providecommand{\U}[1]{\protect\rule{.1in}{.1in}}
\begin{document}
\title{The effect of noise fluctuation of a quantum tunneling device coupled to a substrate}
\author{Nikhilesh A. Vaidya}
\affiliation{Department of Physics, Temple University, Philadelphia, PA 19122, USA}
\author{D. H. Santamore}
\affiliation{Department of Physics and Engineering Physics, Delaware State University,
Dover, DE 19901, USA}

\begin{abstract}
The recent experiment of Stettenheim, et al.~showed that, contrary to
conventional belief, the coupling of a quantum electronic device to its
substrate can have important effects on the noise power spectrum, since the
substrate functions as a mechanical oscillator. We carry out a theoretical
analysis of this coupling in the case of a quantum point contact (QPC). First
we derive the noise power spectrum from the Hamiltonian without making the
Markovian approximation, and obtain numerical results that reproduce the
experimental data. Next we investigate the nature of the coupling. In most
previous analyses, the coupling of an electronic device to a mechanical
oscillator has been modeled as a position coupling. We model it both as a
position coupling and as a momentum coupling and compare the results. We find
that, as long as one includes backaction between position and momentum, the
assumed mode of coupling makes little difference, since the backaction
transmits momentum fluctuations to position fluctuations and vice versa.
Finally, we ask whether the salient features of the model persist in the
Markovian approximation. We find that a Markovian analysis confirms the
QPC-substrate coupling, but underestimates the noise floor and leads to
excessively sharp and narrow noise peaks around the resonant frequencies.

\end{abstract}

\pacs{73.23.Hk, 74.78.Na, 72.10.-d, 05.40.-a}
\keywords{Backaction, Quantum Noise, Fluctuations, Spectral Density, Quantum Point
Contact, Nanoelectromechanical systems, Electron transport}\date{\today }
\maketitle

\section{Introduction}

In recent years, there has been rapid progress in developing new electronic
devices, in which electrical and mechanical degrees of freedom are coupled
\cite{CJG11,STB10,OYN09,SWN07,UGM04,CB04,ZB02,EB01,Graig00,SLN13,Arm04}.
Electron tunneling devices are especially interesting, since they exhibit a
fascinating interplay between macroscopic and mesoscopic phenomena, opening up
new avenues in the study of high precision measurement
\cite{KBV06,AS05,PH05,PG05,PB02,WKR05,GC05,MSS01}. The leading candidates for
ultra-precision motion and mass sensors indlude quantum dots, single electron
transistors, and quantum point contacts (QPCs) \cite{PJR08}.

All devices are built on substrates that provide them with a sturdy and
compact platform, yet hardly any attention has been given to the substrates:
as they are not considered the interesting parts of the system, they have been
ignored in most analyses. However, a recent experiment by Stettenheim et al.
\cite{STB10} revealed a surprising fact: some electronic devices are naturally
coupled to their substrates, which function as mechanical oscillators. In
their experiment, the device-substrate coupling is seen in the power noise
spectrum: the fundamental harmonics of the substrate resonant bending mode
frequency, $f_{\mathrm{b}}$, appeared in the noise spectrum as two spikes $\pm
f_{\mathrm{b}}$\ away from the device circuit resonant frequency. This noise
can reduce the device's sensitivity by decreasing the coherence between the
sample and the device coupling. Since all electronic devices are necessarily
built on substrates, this type of coupling is unavoidable. Therefore, it is
essential to understand the device-substrate coupling mechanism and its effects.

We investigate the coupling mechanism of an electron tunneling device with its
substrate. To demonstrate clearly the effects of substrates, we analyze a
specific type of electronic device -- a QPC -- since experimental noise data
for QPCs are publicly available. This paper has three objectives. (1) We
introduce a model that faithfully reproduces the experimental result of
Ref.\ \cite{STB10}, (2) We resolve a long-standing issue concerning the nature
of the coupling between an electronic device and a mechanical oscillator: is
it a position coupling or a momentum coupling? Most previous work has assumed
that the coupling is a position coupling, but we show that, under certain
conditions, the assumption of momentum coupling gives equally good results.
(3) We examine the applicability of the Markovian approximation that is
frequently used for electronic device analyses.

In connection with the first objective, there have been some previous attempts
to reproduce the noise spectrum \cite{BB12}. These studies have successfully
reproduced the location of the fundamental harmonic peaks around the circuit
resonant frequency, but they do not capture the magnitude of the peaks and the
noise floor seen in the experimental data. Our model improves on these
previous work by incorporating backaction, evaluating the dynamics without
relying on the Markovian approximation, and utilizing parallel computing. We
derive the quantum master equation from the Hamiltonian. Then we calculate the
correlations between the oscillator and the detector, analytically derive the
transport properties of the current, and numerically evaluate the noise
spectrum, which we then compare with the experiment of Stettenheim et al.
\cite{STB10}.

Our second and third objectives have to do with the validity and applicability
of some common assumptions made in modeling electronic devices involving
mechanical oscillators: (a) that the coupling between the system and a
mechanical oscillator is mediated by position, (b) that the Markovian
approximation on the reservoirs is {justified}.

The coupling between an electronic device and a mechanical oscillator is
commonly assumed to be a position coupling, with only a few researchers
positing a momentum coupling. However, there haven't been no firm evidence as
to which coupling assumption is appropriate or whether both are valid. Note
that we are concerned with harmonic coupling only for a position coupling.
Anharmonic coupling either does not exist or is negligible in this context. We
address this long-standing puzzle in terms of fluctuation noise. We examine
the two coupling assumptions separately, and calculate the dynamics and noise
spectrum for each coupling. We show that backaction is the key to the coupling
argument. In position coupling, momentum \textquotedblleft kicks
back\textquotedblright\ electrons that tunnel through the junction, which then
transmit back the \textquotedblleft kick" effect to the oscillator position,
creating a feedback loop between position and momentum. In momentum coupling,
the roles of position and momentum are exactly reversed. As a result, once a
feedback loop is established, both position and momentum fluctuations equally
affects the system through backaction regardless of which coupling one starts
with. We argue that as long as one includes the backaction, the type of
coupling does not materially affect the system's steady-state noise spectrum.

Finally, we examine whether the Markovian approximation is sufficient to
capture all the crucial features in the noise spectrum. We analytically solve
and evaluate the dynamics in the Markovian approximation, and then compare the
results with those obtained from full non-Markovian calculations. We show that
the Markovian approximation can capture the oscillator's fundamental bending
mode coupling signature of noise peaks at the oscillator resonance frequency,
but substantially underestimates the noise floor and predicts sharper noise
peaks than those seen in the full numerical non-Markovian model.

\section{Model and derivations\label{model}}

\begin{figure}[pth]
\begin{center}
\includegraphics[width=\columnwidth] {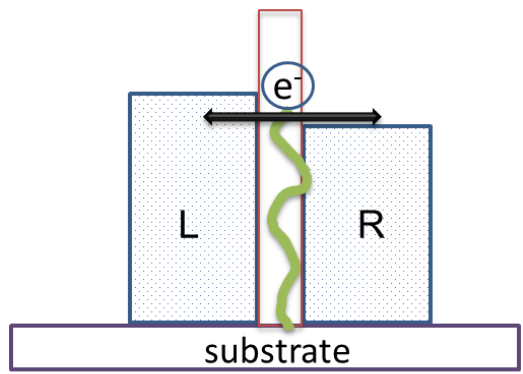}
\end{center}
\caption{Schematics of the model. Two reservoirs (right and left) are
separated by a barrier. The whole system is built on a substrate that acts as
a mechanical oscillator. The oscillator's bending mode is coupled to the
tunneling electrons. A bias is applied to cause electrons to tunnel from the
left reservoir to the right reservoir (forward bias).}%
\label{fig1_model}%
\end{figure}

Our model consists of a QPC and a substrate that acts as a mechanical
oscillator. We choose a QPC as our representative electron tunneling device
for the sake of definiteness and clarity, and to compare the noise spectrum
with that of the experiment in Ref.\ \cite{STB10}. However, our method can be
applied to any tunneling junction device. The QPC has two fermionic
reservoirs, noted as left (L) and right (R). We assume that the tunneling
electrons couple to the substrate oscillator's fundamental bending mode by
position coupling. Figure \ref{fig1_model} shows a schematic of our model. The
Hamiltonian of the system is%
\begin{equation}
H_{tot}=H_{osc}+H_{res}+{H}_{int}, \label{Hamiltonian}%
\end{equation}
where $H_{osc}$ and $H_{res}$ are the non-interacting Hamiltonians for the
substrate and the electron reservoirs,%
\begin{equation}
H_{osc}=\frac{p^{2}}{2M}+\frac{1}{2}M\omega_{0}^{2}x^{2}, \label{H_osc}%
\end{equation}%
\begin{equation}
H_{res}=\sum_{k}\left(  \varepsilon_{L,k}a_{L,k}^{\dag}a_{L,k}+\varepsilon
_{R,k}a_{R,k}^{\dag}a_{R,k}\right)  . \label{H_res}%
\end{equation}
Here, $\omega_{0}$ is the oscillator resonant frequency, $M$ is the mass of
the substrate oscillator and $x$ and $p$ are the position and momentum
operators of the oscillator, $\varepsilon_{L\left(  R\right)  ,k}$ is the
energy of the electrons in the left (right) reservoir with momentum $k$, and
$a\left(  a^{\dag}\right)  $ is the electron annihilation (creation) operator.

The interaction Hamiltonian ${H}_{int}$ describes the tunneling electrons that
are coupled to the substrate, and is given by%
\begin{equation}
{H}_{int}=T(x;t)\sum_{k,q}{Y}^{\dagger}a_{R,k}^{\dag}a_{L,q}+H.c., \label{Hc}%
\end{equation}
where $T(x;t)$ is the electron tunneling amplitude matrix, which depends on
the oscillator's position and includes the backaction, the density of states
of electrons in the reservoirs, tunnelling amplitude coefficients for with and
without the oscillator coupling, and the phase difference between the
non-coupled and coupled tunneling amplitudes \cite{CG04}. To obtain the charge
transport and the charge transfer statistics, we employ a charge counting
method originally developed by Shelankov and Rammer
\cite{Clerk04b,Rammer04,WKR05,BC08}. This method extracts particle transfer
information directly from the wave function of a many-body system. The charge
counting operators\ $\hat{Y}$, $\hat{Y}^{\dagger}$\ are dimensionless and
contain charge projection operators that project the state of the conduction
electrons onto the density matrix of the electrons. These operators count the
number of electron that have tunnelled from the left reservoir to the right
reservoir. Details of the method and the full expression of $T\left(
x;t\right)  $ and $\hat{Y}$ are in Appx.\ \ref{sec AppendixA}.

The total density matrix $\rho_{tot}$ contains all information about the
system and the oscillator and their interactions. The oscillator density
matrix $\rho_{osc}$ evolution is described by the equation of motion%
\begin{equation}
\frac{d}{dt}\rho_{osc}\left(  t\right)  =\frac{1}{i\hbar}\left[
H_{tot}\left(  t\right)  ,\rho_{tot}\left(  t\right)  \right]  .
\end{equation}
We assume that (a) the coupling between the oscillator and the tunneling
electrons is weak, and (b) the density matrices $\rho_{osc}$ and $\rho_{res}$
of the oscillator and the electron reservoirs are initially uncorrelated,
$\rho\left(  t=0\right)  =\rho_{osc}\left(  t=0\right)  \otimes\rho
_{res}\left(  t=0\right)  $. Then, we can use the Born approximation to obtain
the reduced master equation of the oscillator%
\begin{align}
\frac{d}{dt}\rho_{osc}\left(  t\right)   &  =-\frac{1}{\hbar^{2}}%
\int\limits_{0}^{t}dt^{\prime}\nonumber\\
&  \times\mathrm{Tr}_{res}\left[  H_{int}\left(  t\right)  ,\left[
H_{int}\left(  t^{\prime}\right)  ,\rho_{osc}\left(  t\right)  \otimes
\rho_{res}\right]  \right]  . \label{mastereqn}%
\end{align}
To solve the master equation and find the noise spectrum, we take a similar
approach to that of Doiron \cite{DTB07} and calculate an equation of motion
for the oscillator's $n$-resolved density matrix. We do not adopt the
Markovian approximation or any other simplifications. The Markovian
approximation is useful when the two-time correlations of the reservoir decay
much faster than the coherence time between the QPC and the substrate
oscillator in the system, so that we can substitute $\rho_{osc}\left(
t,t^{\prime}\right)  \rightarrow\rho_{osc}\left(  t\right)  $ and replace the
upper time limit $t$ in the integral by $\infty$ \cite{Breuer10}. However,
since we do not know the rate of the correlation decay beforehand, we perform
the un-assumed, full numerical integration, and then, compare the results with
those obtained in the Markovian approximation. We show in
Sec.\ \ref{sub Markov} that the Markovian approximation misses some important
features of the noise spectrum.

From Eq.\ (\ref{mastereqn}), we obtain the number-resolved ($n$-resolved)
master equation%
\begin{equation}
\frac{d}{dt}\rho_{osc}(n,t)=-\frac{1}{\hbar^{2}}\sum_{k,q}\frac{1}{\Lambda
}\int_{0}^{t}dt^{\prime}~\left[  U\left(  n\right)  +V\left(  n\right)
\right]  . \label{n-resolve}%
\end{equation}
Here $n$ is the number of charges transferred, $\Lambda$ is the density of
states for the electrons in the reservoirs originally in the tunneling
amplitude matrix $T$, and $U\left(  n\right)  +V\left(  n\right)  $ contains
the tunneling counting information and the two-time reservoir correlation
functions, and also depends on the energy of tunneling junction of the QPC
(see Appx.\ \ref{sec AppendixA} for details.) Directly solving
Eq.\ (\ref{n-resolve}) for all $n$ is numerically impractical as $n$ tends to
be large. To get around this computational problem, we introduce the counting
field $\chi$ to change the sum of number $n$ to a field $\rho_{osc}\left(
\chi;t\right)  =\sum_{n}e^{i\chi n}\rho_{osc}\left(  n;t\right)  $ that
describes charge transfer events. After some manipulations, we obtain the
unconditional master equation%
\begin{equation}
\frac{d}{dt}\rho_{osc}\left(  t\right)  =\varrho_{0}\left(  \chi=0,t\right)
+\varrho_{1}\left(  \chi,t\right)  , \label{master equation}%
\end{equation}
where $\varrho_{0}$ and $\varrho_{1}$ contain all correlations and transport
properties (c.f.\ Appx.\ \ref{sec AppendixA}).

We can now calculate the transport properties for the forward bias regime. The
average current is given by $\left\langle I\left(  t\right)  \right\rangle
=\left(  2e\right)  d\left\langle n\left(  t\right)  \right\rangle \left/
dt\right.  $, where $e$ is the electron charge, $n$ is the number of
transferred electrons across the junction at time $t$, and $\left\langle
n\right\rangle $ is given by%
\begin{equation}
\frac{d}{dt}\left\langle n\left(  t\right)  \right\rangle =i\mathrm{Tr}%
_{osc}\left[  \frac{d}{d\chi}\left(  \frac{d}{dt}\rho_{osc}\left(
\chi;t\right)  \right)  \right]  _{\chi=0},
\end{equation}
with $d\rho_{osc}\left/  dt\right.  $ is given by Eq.\ (\ref{master equation}%
). We first take the derivative of Eq.\ (\ref{master equation}) with respect
to the counting field conjugate to the transferred charge $n.$ Performing the
calculations and simplifying, we obtain%
\begin{equation}
\frac{\left\langle I\left(  t\right)  \right\rangle }{2e}=\frac{1}{~2e}\left(
\left\langle I\right\rangle _{0}+\left\langle I\right\rangle _{x}+\left\langle
I\right\rangle _{p}+\left\langle I\right\rangle _{xp}+\left\langle
I\right\rangle _{q}\right)  . \label{<I(t)>/2e}%
\end{equation}
Here $\left\langle I\right\rangle _{0}$ is the current without the oscillator
coupling, $\left\langle I\right\rangle _{x}$ and $\left\langle I\right\rangle
_{p}$ are the currents modulated by the coupled oscillator through the
oscillator's position and through momentum, $\left\langle I\right\rangle
_{xp}$ contains the backaction channel that connects position and momentum,
and $\left\langle I\right\rangle _{q}\ $is the quantum correction to the
overall current. The analytical expressions of each term are in
Appx.\ \ref{AppSec_current}.

Next we determine the spectral density $S(\omega)$\ of the current by
calculating the variance of $n(t)$. The current noise spectrum is given by%
\begin{equation}
S\left(  \omega\right)  =\int_{-\infty}^{\infty}dte^{i\omega t}\left\langle
\left\{  \delta I\left(  t\right)  ,\delta I\left(  0\right)  \right\}
\right\rangle , \label{noise}%
\end{equation}
where $\delta I$ is the fluctuation of the current. A coupled system
eventually loses its coherence through interactions with the environment. If
the coherence time of the QPC and the substrate oscillator is longer than the
decay time of the correlation function of the electrons in the right and left
reservoirs, then one can use the MacDonald formula to analytically calculate
the spectral density \cite{M49}.\ However, if the reservoir electron's
correlation decays much more slowly, the MacDonald formula is not applicable.
We perform a full numerical evaluation of the modified spectral density
\cite{WT11} given by%
\begin{equation}
\left\langle S\left(  \omega\right)  \right\rangle =2e^{2}\omega\int%
_{0}^{t^{\prime}}dte^{i\omega t}\frac{d}{dt}\left\langle \left\langle
n^{2}\left(  t\right)  \right\rangle \right\rangle ,
\end{equation}
where $\left\langle \left\langle .\right\rangle \right\rangle $\ denotes
covariance and $\left\langle \left\langle n^{2}\left(  t\right)  \right\rangle
\right\rangle $ is obtained by
\begin{equation}
\frac{d}{dt}\left\langle \left\langle n^{2}\left(  t\right)  \right\rangle
\right\rangle =\frac{d}{dt}\left\langle n^{\text{ }2}\left(  t\right)
\right\rangle -2\left\langle n\left(  t\right)  \right\rangle \frac{d}%
{dt}\left\langle n\left(  t\right)  \right\rangle .
\end{equation}
All higher moments and the correlations between the oscillator and the
transferred charge $n$ are calculated from Eqs.\ (\ref{master equation}) and
(\ref{Moments}). Performing the calculations and simplifying, we obtain%
\begin{align}
\frac{d}{dt}\left\langle \left\langle n^{2}\left(  t\right)  \right\rangle
\right\rangle  &  =\frac{d}{dt}\left\langle n\left(  t\right)  \right\rangle
+\tilde{N}_{\mathrm{x}}\left(  x,t\right)  +\tilde{N}_{\mathrm{p}}\left(
p,t\right) \nonumber\\
&  +\tilde{N}_{\mathrm{xp}}\left(  x,p,t\right)  , \label{d<n^2(t)>/dt}%
\end{align}
where $\tilde{N}_{\mathrm{x}}\left(  x,t\right)  $, $\tilde{N}_{\mathrm{p}%
}\left(  p,t\right)  $, and $\tilde{N}_{\mathrm{xp}}\left(  x,p,t\right)  $
are expressions for the noise associated with the correlations $\left\langle
\left\langle xn\right\rangle \right\rangle $, $\left\langle \left\langle
pn\right\rangle \right\rangle $, and $\left\langle \left\langle \left\{
x,p\right\}  n\right\rangle \right\rangle $, respectively as well as with
their higher moments. The details of these terms are in
Appx.\ \ref{AppSec_current_noise}. Solving the differential equations for the
correlations turns out to be very computationally demanding. Therefore, we use
the double exponential oscillatory method developed by Takahasi and Mori
\cite{TM74}, in which the variable $\omega$ is transformed and the trapezoidal
rule is used to solve the transformed integral instead of the original
integral \cite{OM99}.

\section{Results and discussion}

\subsection{Comparison of theory and experiment\label{sub th vs exp}}

In this section, noise power spectrum $S\left(  \omega\right)  $ is converted
to $P_{n}\left(  \mathrm{dBm}\right)  $ to compare to the experimental results
in Ref.\ \cite{STB10} by first calculate $P_{n}\left(
W\right)  $ in Watts:%
\begin{equation}
P_{n}\left(  W\right)  =S\left(  \omega\right)  e^{2}\omega_{0}\left(
\frac{2L}{CZ}\right)  B_{W}%
\end{equation}
and then, converted to a millidecibel scale. Here $C=0.28$ $\mathrm{pF}$ is
the capacitance and $L=140$ $\mathrm{nH}$ is the inductance, and $Z=50$ is the
impedance of the $LC$ tank circuit. $B_{w}$ is the bandwidth from the measurement.

\begin{figure}[pth]
\begin{center}
\includegraphics[width=\columnwidth] {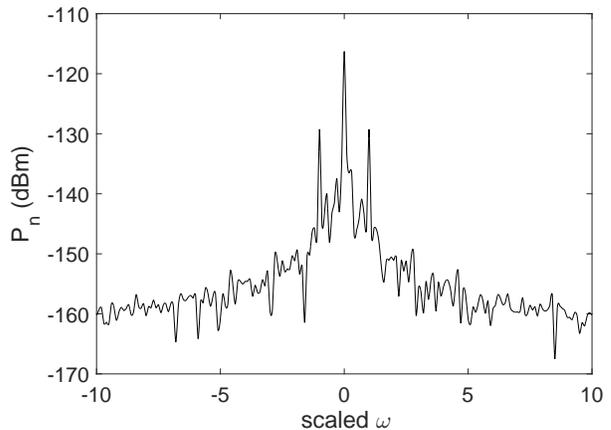}
\end{center}
\caption{Noise power spectrum as a function of dimensionless frequency. The
temperature $90$\textrm{mK}, circuit resonant frequency $f_{0}=800$%
\textrm{MHz}, applied voltage $V=1$\textrm{mV}, tunneling amplitudes
$\tilde{\tau}_{0}=$ $4\times10^{-2}$ $\ $and $\tilde{\tau}_{1}=5\times10^{-5}%
$, and phase shift $\eta=\pi/2$\ are chosen to simulate the experiment in
Ref.\ \cite{STB10}.}%
\label{fig2_Noise}%
\end{figure}

Figure \ref{fig2_Noise} shows the noise power spectrum $P_{n}$\ as a function
of the dimensionless frequency $\tilde{\omega}$, with the parameters used in
the experiments in Ref.\ \cite{STB10}. In the experiment, the QPC was embedded
in an LC tank circuit. The resonant frequency of the LC circuit was
$f_{0}=800$ $\mathrm{MHz}$, the electron reservoir temperature was
$T=90~\mathrm{mK}$, and the bias voltage was $V=1~\mathrm{mV}$.

We set our frequency scale so that $\tilde{\omega}=0$ corresponds to $f_{0}$
and $\tilde{\omega}=\pm1$ corresponds to the fundamental harmonic frequency of
the substrate oscillator bending mode ($f_{\mathrm{b}}=580$ $\mathrm{kHz}$),
with subsequent integers corresponding to higher harmonics. In the experiment,
the forward bias which causes electrons to tunnel from the left reservoir to
the right reservoir is large enough that electron tunneling in the opposite
direction, from right to left, is insignificant. Our calculations also confirm
that the right-to-left tunneling amplitudes is nearly zero.

The theoretical noise calculation matches the experimental data (Figure 2(c)
in Ref.\ \cite{STB10}) very well. The calculation reproduces the correct
position of the noise peaks and the magnitudes of the noise floor. The peak at
$\tilde{\omega}=0$ is due to the fluctuation of the electron current
$\left\langle n\right\rangle $ in the LC circuit. The peaks at $\tilde{\omega
}=\pm1$ indicate that the QPC is coupled to the fundamental harmonic of the
bending mode of the oscillator. Our analysis of each current fluctuation term
in Eq.\ (\ref{d<n^2(t)>/dt}) reveals that the major contribution to the
coupling noise comes from the noise terms $\tilde{N}_{\mathrm{x}}\left(
x,t\right)  $ and $\tilde{N}_{\mathrm{p}}\left(  p,t\right)  $, which are
associated with the position and momentum fluctuations, respectively, and
their contributions are comparable.

The position and momentum fluctuations are connected through backaction via
the correlation $\left\langle \left\langle \left\{  x,p\right\}
n\right\rangle \right\rangle $ contained in $\tilde{N}_{\mathrm{xp}}\left(
x,p,t\right)  $. Although the fluctuation amplitude of $\tilde{N}%
_{\mathrm{xp}}\left(  x,p,t\right)  $ is only about $1/1000$ as large as the
position and momentum fluctuation amplitudes, the backaction itself has a
large role in transmitting the \textquotedblleft kicks\textquotedblright%
\ between the position and momentum fluctuations, amplifying both fluctuation
amplitudes. We can see this quantitatively: if the backaction is absent
($\tilde{N}_{\mathrm{xp}}\left(  x,p,t\right)  =0$), both $\tilde
{N}_{\mathrm{x}}\left(  x,t\right)  $ and $\tilde{N}_{\mathrm{p}}\left(
p,t\right)  $ dramatically decrease and the overall noise floor becomes $20$
\textrm{dBm} lower than it would be with $\tilde{N}_{\mathrm{xp}}\left(
x,p,t\right)  $ present.

\subsection{Position or momentum coupling?\label{sub x or p couple}}

The central role of the backaction raises the question whether the QPC is
coupled to the substrate oscillator through position coupling or momentum
coupling. Most researchers take the coupling between electronic devices and
mechanical oscillators to be position-based, and we have adopted this
assumption up to now in our analysis of the QPC-substrate coupling.

The dramatic decrease in both position and momentum fluctuations in the
absence of backaction as seen in Sec.\ \ref{sub th vs exp} indicates that the
backaction term acts as a channel between position and momentum, and transmits
every kick\ (backaction) from one to the other. Therefore, we now hypothesize
that the type of coupling ---position or momentum--- will not affect the total
amount of noise fluctuations, provided the backaction is included in the
calculations and the system is given sufficient time to respond to the
position and momentum's mutual kicks, so that both position and momentum
influence is fully incorporated into the system dynamics.

\begin{figure}[pth]
\begin{center}
\includegraphics[width=\columnwidth] {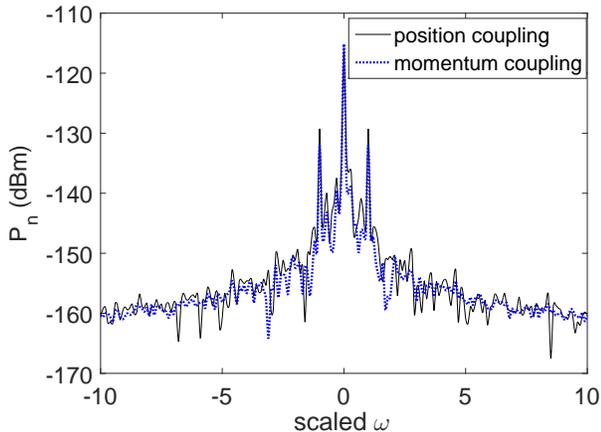}
\end{center}
\caption{Power spectrum of the position coupling (blue dotted line) and the
momentum coupling (black solid line) as a function of dimensionless frequency.
The parameters used are the same as those in Fig.\ \ref{fig2_Noise}.}%
\label{fig3_pos-mom}%
\end{figure}

To examine this hypothesis, we change the position coupling contained in
$T(x;t)$ to momentum coupling by replacing $x/x_{0}$ with $p/p_{0}$ in
Eq.\ (\ref{T(x)}) in Appx.\ \ref{sec AppendixA}. Then, we re-derive the
equation of motion and expressions for the currents, and evaluate the power
noise spectrum. The resulting noise spectra are shown in
Fig.\ \ref{fig3_pos-mom}. The peak noise heights, the widths of the peaks, and
the magnitudes of noise floors are almost identical for both position and
momentum coupling. This result highlights again the importance of backaction
and shows the two couplings to be equivalent in terms of observable
fluctuation noise. Note that our analyses pertain only to noise fluctuations.
We make no claims about the equivalence of the two couplings schemes with
respect to other transport quantities or the transient dynamics. We emphasize
again that the position coupling in our model is a harmonic coupling.
Anharmonic coupling, while potentially interesting, is beyond the scope of
this paper.

\subsection{Information loss in the Markovian approximation\label{sub Markov}}

Finally, we examine the applicability of the Markovian approximation. As
stated before, the Markovian approximation assumes that the system has a short
memory meaning that the correlation time of electrons in the reservoirs is
shorter than the coherence time between the substrate oscillator and the
electrons. On this assumption, many time-dependent tunneling coefficient terms
(such as $\xi$, $D$, $\gamma$, in Appendix \ref{sec AppendixA}) become zero or
negligible. These simplifications make the tunneling parameters
time-independent, and enables the equations to be solved analytically.
However, in the Markovian approximation, $\tilde{N}_{\mathrm{xp}}$ is almost
zero ($\tilde{N}_{\mathrm{xp}}$ is a factor of $10^{-10}$ smaller than that of
non-Markovian) even if the backaction term is included. As a result, the
Markov approximation largely underestimates the noise in all terms of
Eq.\ (\ref{d<n^2(t)>/dt}).

\begin{figure}[pth]
\begin{center}
\includegraphics[width=\columnwidth] {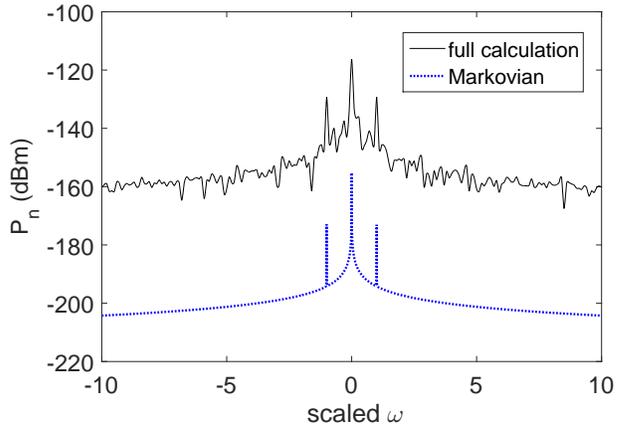}
\end{center}
\caption{Noise power spectrum as a function of dimensionless frequency.
Comparing the Markov approximation (blue dotted line) to the full
non-Markovian approxiamaton (black solid line). The parameters used are the
same as those in Fig.\ \ref{fig2_Noise}.}%
\label{fig4_Markov}%
\end{figure}

Figure \ref{fig4_Markov} shows the power noise spectra for both the full
non-Markovian evaluation and the Markovian approximation. The non-Markovian
calculation is done numerically, whole the Markovian calculation is done
analytically. Both spectra show the LC circuit resonant frequency and the
first harmonic of the substrate oscillator bending mode coupling as noise
peaks. On the other hand, the amplitude of the noise floor in the Markovian
approximation is about $45$ \textrm{dBm} lower than in the full non-Markovian calculation.

Part of this gap comes from backaction, which is negligible in the Markovian
approximation and, as shown in Sec.\ \ref{sub th vs exp}, contributes about
$20$ \textrm{dBm} to the noise floor. Other factors such as shot noise also
contribute to the noise floor, which the Markovian approximation again
severely underestimates by another $\sim25$ \textrm{dBm}.

Another difference between the Markovian and non-Markovian calculations is
that the noise peaks in the Markovian approximation are very sharp spikes,
whereas in the full non-Markovian calculation, the peaks are not as sharp and
tall, but rather are broadened around the resonant frequencies.

\begin{figure}[pth]
\begin{center}
\includegraphics[width=\columnwidth] {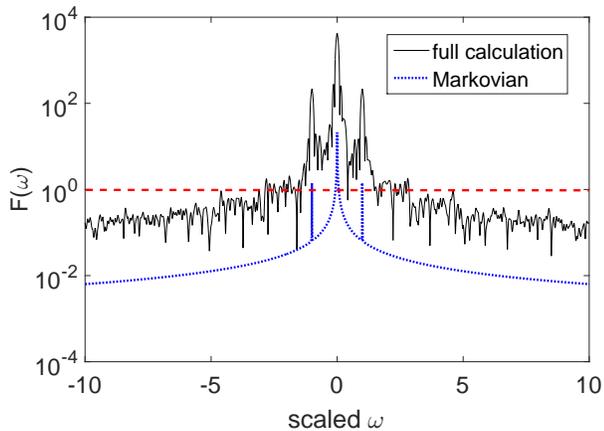}
\end{center}
\caption{Fano factor as a function of scaled frequency for both Markovian
(blue dotted line) and non-Markovian approxiamaton (black solid line) cases.
The parmeters used are the same as those in Fig.\ \ref{fig4_Markov}. $F=1$
(red dashed line) corresponds to a Poisson process. $F>1000$ is regarded as
super-Poisson.}%
\label{fig5_FanoMarkov}%
\end{figure}

The Fano factor is the ratio of total noise to current noise as a function of
the dimensionless frequency. Physically, the Fano factor is a good indicator
of the strength of the coupling between the electrons and the oscillator.
Figure \ref{fig5_FanoMarkov} shows Fano factors for both the Markovian
approximation and full non-Markovian calculations. The noise reaches the
super-Poisson range ($>1000$) in both the experiment and full calculation, but
not in the Markovian approximation. This, too, contributes to the Markovian
approximation's underestimating overall noise.

We see, then, that a simple Markovian approximation calculation suffices to
demonstrate the existence and significance of the coupling of tunneling
devices to substrates, since the noise peaks due to this coupling are clearly
visible even in the Markovian plot. However, if one theoretically tries to
explore a possibility to reduce the noise to increase the sensitivity of a QPC
sensor, or to analyze a new device that takes an advantage of this extra
mechanical degree of freedom in the future, then, the full numerical
non-Markovian calculation is necessary.

\section{Conclusion}

We have modeled electron transport and its noise spectrum for a tunnel
junction device coupled to a substrate that acts as a mechanical oscillator.
We focused on a QPC, so as to compare with available experimental data. In the
first place, we made the standard assumption that the QPC and the substrate
are coupled through position. In addition, we included backaction in our
model. We obtained the noise spectrum from a full numerical evaluation,
without making the Markovian approximation. We found that the current noise is
strongly modified by the mechanical degrees of freedom. There are sharp peaks
around the fundamental harmonic frequency of the substrate bending mode, and
the noise floor is increased. Our results reproduce all key features of the
experimental results of Ref.\ \cite{STB10}. Next we examined the nature of
coupling. We calculated the noise spectrum for both position coupling and
momentum coupling and found that the choice of coupling scheme does not make a
difference in the noise spectrum as long as backaction is included and the
noise spectra are evaluated in the steady state. The backaction acts as a
channel between position and momentum influencing each other and amplify the
fluctuations. Finally, we investigated the validity of the Markovian
approximation. We compared the Markovian approximation with the full numerical
result. While the Markovian approximation still shows key features, such as
the fundamental resonance mode noise spike, it loses some important features,
such as super-Poisson noise, noise peak broadening, and the noise floor. The
Markovian approximation underestimate the effect of backaction, as well as
other quantum noise, such as shot noise.

\begin{acknowledgments}
We thank Andreas Metz and Jonathan Tannenhauser for useful discussions.
\end{acknowledgments}

\appendix

\section{$n-$resolved master equation\label{sec AppendixA}}

The Hamiltonian of the system is%
\begin{equation}
H_{tot}=H_{osc}+H_{res}+{H}_{int}, \label{App_Hamiltonian}%
\end{equation}
where $H_{osc}$ and $H_{res}$ are the non-interacting Hamiltonians for the
substrate and the electron reservoirs. The objective here is to derive the
master equation Eq.\ (\ref{master equation}) in the main text.

We use the interaction picture and which represents our specific system from
the general, time local, non-Markovian master equation Eq. (\ref{mastereqn}).
\
\begin{align}
\frac{d}{dt}\rho_{osc}(t) &  =-\frac{1}{\hbar^{2}}\int\limits_{0}%
^{t}dt^{\prime}Tr_{res}\left[  H_{int}(t),\right.  \nonumber\\
&  \left.  \left[  H_{int}(t^{\prime}),\rho_{osc}(t)\otimes\rho_{res}\right]
\right]  \text{ },\label{masterAppx}%
\end{align}
where the interaction Hamiltonian, ${H}_{int}$, describes the tunneling
electrons that are coupled to the substrate, and is given by
\begin{equation}
{H}_{int}=T\left(  x;t\right)  \sum_{k,q}{Y}^{\dagger}a_{R,k}^{\dag}%
a_{L,q}+H.c..\label{HcAppx}%
\end{equation}
We employ a charge counting method mentioned in the main text, which count the
tunneling electrons $n$ from the left reservoir to the right reservoir through
the tunneling junction. The dimensionless charge counting operators $\hat{Y}$
and $\hat{Y}^{\dagger}$ are defined as%
\begin{align}
\hat{Y}\rho\left(  n;t\right)   &  \equiv\rho\left(  n+1;t\right)  \left(
t\right)  ,\\
\hat{Y}^{\dagger}\rho\left(  n;t\right)   &  \equiv\rho\left(  n-1;t\right)
\left(  t\right)  .
\end{align}
The oscillator-position dependent tunneling amplitude of electrons $T\left(
x;t\right)  $ is
\begin{equation}
T(x,t)=\frac{1}{\Lambda}\left(  \tilde{\tau}_{0}+e^{i\text{ }\eta}\tilde{\tau
}_{1}\frac{x\left(  t\right)  }{x_{0}}\right)  ,\label{T(x)}%
\end{equation}
and%
\begin{equation}
x\left(  t\right)  =x\cos\omega_{0}t+\frac{p}{M\omega_{0}}\sin\omega
_{0}t,\label{xInInteractionPic}%
\end{equation}
where $\Lambda=\Lambda(\epsilon_{L,q},\epsilon_{R,k})$ is the density of
states and a function of the energy of the electrons in the left,
$\epsilon_{L,q}$, and right $\epsilon_{R,k}$, reservoirs. $\tilde{\tau}_{0}$
and $\tilde{\tau}_{1}$\ are the dimensionless tunneling amplitudes without and
with the vibration mode coupling, respectively. $\eta$ is the phase difference
between the $\tilde{\tau}_{0}$ and $\tilde{\tau}_{1},$ and $x_{0}=\sqrt
{\hbar/2M\omega_{0}}$, where $M$ and $\omega_{0}$ are the mass and the
resonant frequency of the substrate oscillator, respectively. A rough value of
$\tilde{\tau}_{1}$is obtained from the coupling coefficient $\lambda$, which
is geometry and material dependent and calculated in Ref.\ \cite{STB10} using
the experimental values. We take this $\lambda$ and convert to dimensionless
tunneling amplitude $\tilde{\tau}_{1}\approx10^{-5}$. $\tilde{\tau}_{0}$ is
set to $\tilde{\tau}_{0}\approx10^{-2}\sim10^{-1}$ \cite{BB12,CG04}. Then we
finely tune these coefficients numerically.

We note that the tunneling amplitude in Eq.\ (\ref{T(x)}) is linear on the
position. This model is valid for the weak coupling which allows us to use the
linear response theory. We also need the temperature range where the energy
cost to add a charge $e$ is larger than the thermal energy, $k_{BT}$. The
model is applicable to wide variety of electronic devices including tunnel
junctions, superconducting SET, and quantum dots. Smirnov et al \cite{SMH03},
considered an exponential coupling to the position to extend an applicability
to arbitrary voltage applied to the junction and arbitrary temperature of
electrons in leads.

Next, we want to find $\left\langle n\right\vert Q\left\vert n\right\rangle $,
where $Q$ is the integrand in Eq.\ (\ref{masterAppx}). For convenience, we
group the operators $T$ and $Y$ as
\begin{equation}
j\left(  x,t\right)  \equiv T\left(  x,t\right)  {Y}^{\dagger}. \label{j(t)}%
\end{equation}
Inserting Eq.\ (\ref{HcAppx}) into $Q$, and tracing out the reservoir part
result in
\begin{align}
&  \text{ }\left\langle n\right\vert Q\left\vert n\right\rangle \nonumber\\
&  =\left[  j(t)\,j^{\dagger}\left(  t^{\prime}\right)  \rho_{osc}%
(n,t)-j^{\dagger}\left(  t^{\prime}\right)  \rho_{osc}(n+1,t)j\left(
t\right)  \right] \nonumber\\
&  \times f_{R}(1-f_{L})e^{-\frac{i}{\hbar}\left(  \epsilon_{L}-\epsilon
_{R}\right)  \left(  t-t^{\prime}\right)  }\nonumber\\
&  +\left[  \rho_{osc}(n,t)\,j\left(  t^{\prime}\right)  j^{\dagger}\left(
t\right)  -j^{\dagger}\left(  t\right)  \rho_{osc}(n+1,t)j\left(  t^{\prime
}\right)  \right]  \text{ }~\nonumber\\
&  \times f_{R}(1-f_{L})e^{\frac{i}{\hbar}\left(  \epsilon_{L}-\epsilon
_{R}\right)  \left(  t-t^{\prime}\right)  }\nonumber\\
&  +\left[  j^{\dagger}(t)\,j\left(  t^{\prime}\right)  \rho_{osc}%
(n,t)-j\left(  t^{\prime}\right)  \rho_{osc}(n-1,t)j^{\dagger}\left(
t\right)  \right]  ~\nonumber\\
&  \times f_{L}(1-f_{R})e^{\frac{i}{\hbar}\left(  \epsilon_{L}-\epsilon
_{R}\right)  \left(  t-t^{\prime}\right)  }\text{ }\nonumber\\
&  +\left[  \rho_{osc}(n,t)\,j^{\dagger}\left(  t^{\prime}\right)  j\left(
t\right)  -j\left(  t\right)  \rho_{osc}(n-1,t)j^{\dagger}\left(  t^{\prime
}\right)  \right]  \text{ }\nonumber\\
&  \times f_{L}(1-f_{R})e^{-\frac{i}{\hbar}\left(  \epsilon_{L}-\epsilon
_{R}\right)  \left(  t-t^{\prime}\right)  }. \label{Eq.11}%
\end{align}
As the reservoirs are in equilibrium, we have the following electron
correlation functions%
\begin{equation}%
\begin{array}
[c]{c}%
\left\langle a_{L}^{\dagger}\left(  t\right)  a_{R}\left(  t\right)
\right\rangle \ =0,\\
\left\langle a_{L}^{\dagger}(t)a_{L^{\prime}}\left(  t^{\prime}\right)
\right\rangle \ =f_{LL^{\prime}}\ \delta_{LL^{\prime}},\\
\left\langle a_{L(R)}^{\dagger}\left(  t^{\prime}\right)  a_{L(R)}\left(
t\right)  \right\rangle =f_{L(R)\text{ }}-\ 1\\
\left\langle a_{\alpha}^{\dagger}a_{\gamma}a_{\beta}^{\dagger}a_{\delta
}\right\rangle =\left\langle a_{\alpha}^{\dagger}a_{\gamma}\right\rangle
\left\langle a_{\beta}^{\dag}a_{\delta}\right\rangle -\left\langle a_{\alpha
}^{\dagger}a_{\delta}\right\rangle \left\langle a_{\beta}^{\dagger}a_{\gamma
}\right\rangle ,
\end{array}
\end{equation}
with $\alpha,\beta,\gamma,\delta=L,R$ and The last equation comes from the
Wick's theorem. $f_{L\left(  R\right)  }$\ is the Fermi distribution function
for the left (right) electron reservoir:%
\begin{equation}
f_{L(R)\text{ }}=\frac{1}{\exp\left[  \left(  \epsilon_{L\left(  R\right)
}-\mu_{L(R)}\right)  /k_{B}T\right]  +1}, \label{FermiFn}%
\end{equation}
where $\mu_{L\left(  R\right)  }$ is the chemical potential of the left
(right) reservoirs, $k_{B}$ is the Boltzmann constant and $T$ is the reservoir temperature.

Equation\ (\ref{Eq.11}) is unfortunately not useful for actual evaluations
since solving for all $n$ numerically is impractical due to $n$ being large.
Thus, we take the counting field approach that transforms the sum of the
number to a continuous field \cite{DTB07}. The number $n$ is transformed to
the counting field $\chi$ as%
\begin{equation}
\rho_{osc}\left(  \chi;t\right)  =\sum_{n}e^{i\chi n}\rho_{osc}\left(
n;t\right)  , \label{IFTof Rho}%
\end{equation}
and its inverse-transform is%
\begin{equation}
\sum_{n}e^{i\chi n}\rho_{osc}(n+\sigma,t)=e^{-i\chi\sigma}\rho_{osc}\left(
\chi;t\right)  , \label{FTofRho}%
\end{equation}
where $\rho_{osc}(\chi,t)$ is the characteristic function describing the
charge transfer events and $\sigma=\pm1$. We define that $\sigma=+1$ means the
charge transfer from the left to the right reservoir (forward bias) and
$\sigma=-1$ from the right to the left reservoir (backward bias).

Transforming Eqs.\ (\ref{masterAppx}) and (\ref{Eq.11}) using
Eq.\ (\ref{IFTof Rho}) and regrouping terms result in the counting field
equivalent of the number-resolved ($n$-resolved) master equation%
\begin{equation}
\frac{d}{dt}\rho_{osc}\left(  \chi;t\right)  =-\frac{1}{\hbar^{2}}\sum
_{k,q}\frac{1}{\Lambda^{2}}\int_{0}^{t}dt^{^{\prime}}\left[  U\left(
\chi,t,t^{\prime}\right)  +V\left(  \chi,t,t^{\prime}\right)  \right]  ,\text{
} \label{App_n-resolve}%
\end{equation}
where $\Lambda$ is the density of states for the electrons in the reservoirs,
which was originally contained in the tunneling amplitude matrix $T$. The
functions $U$ and $V$ contain the tunneling counting information and the
two-time reservoir correlation functions. They also depend on the energy of
tunneling junction of the QPC. The terms $U\left(  \chi,t,t^{\prime}\right)  $
and $V\left(  \chi,t,t^{\prime}\right)  $ are given by%
\begin{align}
U\left(  \chi,t,t^{\prime}\right)   &  =\left[  j(t)j^{\dagger}(t^{\prime
})\rho_{osc}\left(  \chi\right)  -e^{-i\chi}j^{\dagger}(t^{\prime})\rho
_{osc}(\chi)j\left(  t\right)  \right. \nonumber\\
&  \left.  +\rho_{osc}\left(  \chi\right)  j^{\dagger}(t^{\prime})j\left(
t\right)  -e^{i\chi}j\left(  t\right)  \rho_{osc}\left(  \chi\right)
j^{\dagger}(t^{\prime})\right] \nonumber\\
&  \times F_{s}(t,t^{\prime})\nonumber\\
&  +\left[  j(t)j^{\dagger}(t^{\prime})\rho_{osc}\left(  \chi\right)
-e^{-i\chi}j^{\dagger}(t^{\prime})\rho_{osc}\left(  \chi\right)  j\left(
t\right)  \right. \nonumber\\
&  -\left.  \rho_{osc}\left(  \chi\right)  j^{\dagger}(t^{\prime})j\left(
t\right)  +e^{i\chi}j\left(  t\right)  \rho_{osc}\left(  \chi\right)
j^{\dagger}(t^{\prime})\right] \nonumber\\
&  \times F_{a}(t,t^{\prime}), \label{Eq. 14.1}%
\end{align}
and%

\begin{align}
V\left(  \chi,t,t^{\prime}\right)   &  =\left[  \rho_{osc}\left(  \chi\right)
j(t^{\prime})j^{\dagger}\left(  t\right)  -e^{-i\chi}j^{\dagger}\left(
t\right)  \rho_{osc}\left(  \chi\right)  j(t^{\prime})\right. \nonumber\\
&  \left.  +j^{\dagger}\left(  t\right)  j(t^{\prime})\rho_{osc}\left(
\chi\right)  -e^{i\chi}j(t^{\prime})\rho_{osc}\left(  \chi\right)  j^{\dagger
}\left(  t\right)  \right] \nonumber\\
&  \times F_{s}^{\dagger}(t,t^{\prime})\\
&  +\rho_{osc}\left(  \chi\right)  j(t^{\prime})\,j^{\dagger}\left(  t\right)
-e^{-i\chi}\text{ }j^{\dagger}(t)\rho_{osc}\left(  \chi\right)  j(t^{\prime
})\nonumber\\
&  \left.  -j^{\dagger}(t)j(t^{\prime})\rho_{osc}\left(  \chi\right)
+e^{i\chi}j(t^{\prime})\rho_{osc}\left(  \chi\right)  j^{\dagger}\left(
t\right)  \right] \nonumber\\
&  \times F_{a}^{\dagger}(t,t^{\prime}), \label{Eq. 14.2}%
\end{align}
with $\rho_{osc}=(\chi)\rho_{osc}(\chi,t)$. $F_{s}$ and $F_{a}$ are the
symmetric and anti-symmetric two time reservoir correlation functions,
respectively,
\begin{equation}
F_{s}(t,t^{\prime})=\frac{1}{2}\left[  f_{L}\left(  1-f_{R}\right)
+f_{R}\left(  1-f_{L}\right)  \right]  e^{-\frac{i}{\hbar}\left(  \epsilon
_{L}-\epsilon_{R}\right)  \left(  t-t^{\prime}\right)  }, \label{Fs}%
\end{equation}%
\begin{equation}
F_{a}(t,t^{\prime})=\frac{1}{2}\left[  f_{L}\left(  1-f_{R}\right)
-f_{R}\left(  1-f_{L}\right)  \right]  e^{-\frac{i}{\hbar}\left(  \epsilon
_{L}-\epsilon_{R}\right)  \left(  t-t^{\prime}\right)  }. \label{Fa}%
\end{equation}
Using Eqs. (\ref{App_n-resolve}), (\ref{Eq. 14.1}), and (\ref{Eq. 14.2}), we
regroup, perform some algebraic manipulations, and simplify and finally obtain%
\begin{align}
\frac{d}{dt}\rho_{osc}\left(  \chi;t\right)   &  =-\frac{i}{\hbar}\left[
H_{osc}-\frac{F_{1}x}{x_{0}}-\frac{F_{2}p}{x_{0}M\omega_{0}},\rho_{osc}\left(
\chi;t\right)  \right] \nonumber\\
&  -\frac{1}{\hbar^{2}}\sum_{k,q}\frac{1}{\Lambda^{2}}\int_{0}^{t}dt^{\prime
}\left[  \alpha_{s}+\alpha_{a}+\beta_{s}+\beta_{a}\right]  , \label{Eq. 15}%
\end{align}
where $F_{1}\left(  \eta,t\right)  $ and $F_{2}\left(  \eta,t\right)  $ are
backaction energies given by
\begin{equation}
F_{1(2)}(\eta,t)=2\hbar\sin\eta\sum\limits_{\sigma=\pm1}\tilde{\tau}_{0}%
\tilde{\tau}_{1}\xi_{a,\sigma}^{+(-)}, \label{F12}%
\end{equation}
with $\xi_{a,\sigma}^{+(-)}$ the tunneling parameter with its full expression
is found in Eq.\ (\ref{XIsaPlusMinus}). The sum runs for all the energy modes
available. The terms in the second line describe the dynamics of the
oscillator coupled to the QPC. For brevity, we use short-hand notation
$\rho=\rho_{osc}(\chi,t)$, $\alpha_{s}=\alpha_{s}\left(  \chi,t,t^{\prime
}\right)  ,$ $\alpha_{a}=\alpha_{a}\left(  \chi,t,t^{\prime}\right)  ,$
$\beta_{s}=\beta_{s}\left(  \chi,t,t^{\prime}\right)  $ and $\beta_{a}%
=\beta_{a}\left(  \chi,t,t^{\prime}\right)  $. The subscripts $s$ and
$a$\ denote symmetric and antisymmetric functions, respectively. The full
expression of each term is shown below.%

\begin{align}
\alpha_{s}  &  =\left(  2\rho\tilde{\tau}_{0}^{2}+A_{1}+A_{2}+A_{3}%
+A_{4}+A_{5}\right. \nonumber\\
&  +\left.  A_{6}+A_{7}+A_{8}+A_{9}\right)  F_{s}\\
&  -e^{-i\chi}\left(  A_{10}+A_{11}+A_{12}+A_{13}+A_{14}\right)
F_{s}\nonumber\\
&  -e^{i\chi}\left(  \rho\tilde{\tau}_{0}^{2}+A_{15}+A_{16}+A_{17}%
+A_{18}+A_{19}\right)  F_{s}%
\end{align}%
\begin{align}
\alpha_{a}  &  =\left(  B_{1}+B_{2}+B_{3}+B_{4}+B_{5}+B_{6}+B_{7}\right)
F_{a}\nonumber\\
&  -e^{-i\chi}\left(  B_{8}+B_{9}+B_{10}+B_{11}\right)  F_{a}\nonumber\\
&  +e^{i\chi}\left(  \rho\tilde{\tau}_{0}^{2}+B_{12}+B_{13}+B_{14}%
+B_{15}\right)  F_{a}%
\end{align}%
\begin{align}
\beta_{s}  &  =\left(  2\rho\tilde{\tau}_{0}^{2}+C_{1}+C_{2}+C_{3}+C_{4}%
+C_{5}\right. \nonumber\\
&  +\left.  C_{6}+C_{7}+C_{8}+C_{9}\right)  F_{s}^{\dagger}\\
&  -e^{-i\chi}\left(  \rho\tilde{\tau}_{0}^{2}+C_{10}+C_{11}+C_{12}%
+C_{13}\right)  F_{s}^{\dagger}\nonumber\\
&  -e^{i\chi}\left(  \rho\tilde{\tau}_{0}^{2}+C_{14}+C_{15}+C_{16}%
+C_{17}\right)  F_{s}^{\dagger},
\end{align}%
\begin{align}
\beta_{a}  &  =\left(  D_{1}+D_{2}+D_{3}+D_{4}+D_{5}\right. \nonumber\\
&  +\left.  D_{6}+D_{7}+D_{8}\right)  F_{a}^{\dagger}\\
&  -e^{-i\chi}\left(  \rho\tilde{\tau}_{0}^{2}+D_{9}+D_{10}+D_{11}%
+D_{12}\right)  F_{a}^{\dagger}\nonumber\\
&  +e^{i\chi}\left(  \rho\tilde{\tau}_{0}^{2}+D_{13}+D_{14}+D_{15}%
+D_{16}\right)  F_{a}^{\dagger}.
\end{align}
All terms from $A_{1}$\ to $D_{16}$\ are in Appx.\ \ref{AppSec_coeff_list}.

In the continuous limit, we substitute the sum to an integral over the energy
(thus, over electron frequencies) of the left and the right reservoirs and
write Eq. (\ref{Eq. 15}) as%
\begin{align}
&  \frac{d}{dt}\rho_{osc}(\chi,t)\nonumber\\
&  =-\frac{i}{\hbar}\left[  H_{osc}-F(\eta,t),\rho_{osc}(\chi,t)\right]
\nonumber\\
&  -\int_{0}^{t}dt^{\prime}\int\limits_{0}^{\infty}d\omega_{R,k}%
\int\limits_{0}^{\infty}d\omega_{L,q}\text{ }\left[  \alpha_{s}+\beta
_{s}+\alpha_{a}+\beta_{s}\right]  . \label{Eq. 17}%
\end{align}

Evaluating the integrations in Eq. (\ref{Eq. 17}), we finally obtain the form
of the unconditional master equation that can be numerically evaluated to
obtain the current and noise spectrum:%
\begin{equation}
\frac{d}{dt}\rho_{osc}\left(  t\right)  =\varrho_{0}\left(  \chi=0,t\right)
+\varrho_{1}\left(  \chi,t\right)  , \label{App_master equation}%
\end{equation}
The full expressions for $\varrho_{\text{ }0}$ and $\varrho_{\text{ }1}\left(
\chi,t\right)  $ are:%

\begin{align}
\varrho_{0}  &  =-\frac{i}{\hbar}\left[  H_{osc}-\frac{F_{1}x}{x_{0}}%
-\frac{F_{2}p}{x_{0}M\omega_{0}},\rho\right] \nonumber\\
&  -\sum\limits_{\sigma=\pm1}\left\{  \frac{\tilde{\tau}_{1}^{2}}{x_{0}^{2}%
}\left(  \left[  x,\left[  x,\rho\right]  \right]  D_{s,\sigma}^{+}+\left[
x^{2},\rho\right]  D_{a,\sigma}^{+}\right)  \right. \nonumber\\
&  \left.  +\frac{\tilde{\tau}_{1}^{2}}{x_{0}^{2}M\omega_{0}}\left\{  \left[
x,\left[  p,\rho\right]  \right]  \gamma_{s,\sigma}^{+}+\left[  x,\left\{
p,\rho\right\}  \right]  \gamma_{a,\sigma}^{+}\right\}  \right.  \text{
}\nonumber\\
&  \left.  +\left.  \left[  p,\left\{  x,\rho\right\}  \right]  D_{a,\sigma
}^{-}+\left[  p,\left[  x,\rho\right]  \right]  D_{s,\sigma}^{-}\right\}
\right. \nonumber\\
&  \left.  +\frac{\tilde{\tau}_{1}^{2}\text{ }}{\text{ }x_{0}^{2}M^{2}%
\omega_{0}^{2}}\left(  \left[  p^{2},\rho\right]  \text{\ }\gamma_{a,\sigma
}^{-}+\left[  p,\left[  p,\rho\right]  \right]  \gamma_{s,\sigma}^{-}\right)
\right\}  , \label{newRho0}%
\end{align}%
\begin{align}
\varrho_{1}  &  =-\sum\limits_{\sigma=\pm1}\left(  1-e^{-\text{ }i\text{
}\sigma\chi}\right)  \text{ }\nonumber\\
&  \left.  \left\{  2\tilde{\tau}_{0}^{2}\rho\xi_{s,\sigma}+\frac{\tilde{\tau
}_{0}\tilde{\tau}_{1}}{x_{0}}\left(  x\rho e^{-i\sigma\eta}+\rho
xe^{i\sigma\eta}\right)  \xi_{s,\sigma}^{+}\right.  \right. \nonumber\\
&  \left.  +\frac{\tilde{\tau}_{0}\tilde{\tau}_{1}\text{ }}{x_{0}M\omega_{0}%
}\left(  p\rho e^{-i\sigma\eta}+\rho pe^{i\sigma\eta}\right)  \xi_{s,\sigma
}^{-}\right. \nonumber\\
&  \left.  +\frac{\tilde{\tau}_{0}\tilde{\tau}_{1}}{x_{0}}\left(  x\rho
e^{-i\sigma\eta}-\rho xe^{i\sigma\eta}\right)  \xi_{a,\sigma}^{+}\right.
\nonumber\\
&  \left.  +\frac{\tilde{\tau}_{0}\tilde{\tau}_{1}\text{ }}{x_{0}M\omega_{0}%
}\left(  p\rho e^{-i\sigma\eta}-\rho pe^{i\sigma\eta}\right)  \xi_{a,\sigma
}^{-}\right. \nonumber\\
&  \left.  +\text{ }\frac{\tilde{\tau}_{1}^{2}}{x_{0}^{2}}\left(  x\rho
xD_{s,\sigma}^{+}+p\rho p\frac{\gamma_{s,\sigma}^{-}}{M\text{ }^{2}\omega
_{0}^{2}}\right)  \right. \nonumber\\
&  \left.  -\frac{\tilde{\tau}_{1}^{2}\text{ }}{x_{0}^{2}M\omega_{0}}\left(
x\rho p-p\rho x\right)  \left(  \gamma_{a,\sigma}^{+}-D_{a,\sigma}^{-}\right)
\right. \nonumber\\
&  \left.  +\frac{\tilde{\tau}_{1}^{2}\text{ }}{x_{0}^{2}M\omega_{0}}\left(
\text{~}x\rho p~+~p\rho x\right)  \left(  \gamma_{s,\sigma}^{+}+D_{s,\sigma
}^{-}\right)  \right\}  ,\nonumber\\
&  \text{ } \label{newRho1}%
\end{align}
where $\rho=\rho_{osc}(t)$, $x=x(t)$, $p=p(t)$, and $\xi$, $D$ and $\gamma$
are the tunneling parameters given by%
\begin{equation}
\xi_{s(a),\sigma}(t)=\Gamma_{s(a),\sigma}+\Gamma_{s(a),\sigma}^{~\dagger},
\label{XIsa}%
\end{equation}%
\begin{align}
\xi_{s(a),\sigma}^{~+}(t)  &  =\cos\omega_{0}t~~\xi_{~s(a),\sigma
}(t),~~\nonumber\\
\xi_{~s(a),\sigma}^{~-}(t)  &  =\sin\omega_{0}t~~\xi_{~s(a),\sigma}(t),
\label{XIsaPlusMinus}%
\end{align}%
\begin{align}
D_{a\left(  s\right)  ,\sigma}^{+}(t)  &  =\Gamma_{s(a),\text{ }\sigma}\left(
+\omega_{0}\right)  +\Gamma_{s(a),\text{ }\sigma}\left(  -\omega_{0}\right)
\nonumber\\
&  +\Gamma_{s(a),\text{ }\sigma}^{\dag}\left(  +\omega_{0}\right)
+\Gamma_{s(a),\text{ }\sigma}^{\dag}\left(  -\omega_{0}\right)  ,
\label{DsaPlus}%
\end{align}%
\begin{align}
D_{a\left(  s\right)  ,\sigma}^{-}(t)  &  =-i\left[  \Gamma_{s(a),\text{
}\sigma}\left(  +~\omega_{0}\right)  -\Gamma_{s(a),\text{ }\sigma}\left(
-\omega_{0}\right)  \right] \nonumber\\
&  -i\left[  \Gamma_{s(a),\text{ }\sigma}^{~\dagger}\left(  +~\omega
_{0}\right)  -\Gamma_{s(a),\text{ }\sigma}^{~\dagger}\left(  -\omega
_{0}\right)  \right]  , \label{DsaMinus}%
\end{align}%
\begin{align}
\gamma_{s\left(  a\right)  ,\sigma}^{+}(t)  &  =-iD_{a\left(  s\right)
,\sigma}^{+}\left(  t\right)  ,\nonumber\\
\gamma_{s\left(  a\right)  ,\sigma}^{-}(t)  &  =-iD_{a\left(  s\right)
,\sigma}^{-}\left(  t\right)  , \label{GsaPlusMinus}%
\end{align}
with,
\begin{align}
&  \Gamma_{s(a),\sigma}\left(  \pm~\omega_{0}\right) \nonumber\\
&  =\Gamma_{s(a),\sigma}\left(  \Omega_{L,k}-\Omega_{R,q}+eV/\hbar\pm
\omega_{0},t\right) \\
&  =\frac{1}{2}\int_{0}^{t}dt^{\prime}\int\limits_{0}^{\infty}d\Omega
_{R,\text{ }k}\int\limits_{0}^{\infty}d\Omega_{L,\text{ }q}\mathcal{F}%
_{s(a),\sigma}\nonumber\\
&  \times\exp\left[  i\sigma\left(  \text{ }\omega_{L,k}-\omega_{R,q}\pm
\omega_{0}\right)  \left(  t-t^{\prime}\right)  \right]  ,\nonumber
\end{align}
where we define%
\begin{equation}
\hbar\Omega_{L(R),q(k)}\equiv\hbar\omega_{L\left(  R\right)  ,q\left(
k\right)  }-\mu_{L(R)},
\end{equation}
and $\mu_{L(R)}$ is the chemical potential of the left (right) reservoir, and%
\begin{equation}
\mathcal{F}_{s(a)}=2F_{s(a)}e^{\frac{i}{\hbar}\left(  \epsilon_{L}%
-\epsilon_{R}\right)  \left(  t-t^{\prime}\right)  }.
\end{equation}

With the results above, all the transport properties of the system can be
determined from the Eq. (\ref{App_master equation}) and using the formula%
\begin{align}
&  \frac{d}{dt}\left\langle x^{m_{1}}p^{m_{2}}n^{m_{3}}\left(  t\right)
\right\rangle \nonumber\\
&  \left.  =\left(  i^{m_{3}}\right)  ~Tr_{osc}\left[  x^{m_{1}}p^{m_{2}%
}\right.  \frac{\partial^{m_{3}}}{\partial\chi^{m_{3}}}\frac{d}{dt}\rho
(\chi;t)\right]  _{\chi=0}, \label{Moments}%
\end{align}
where $m_{i}$, is any integer moment number.

\section{Current and noise spectrum}

\subsection{Current\label{AppSec_current}}

The average current is given by
\begin{equation}
\left\langle I\left(  t\right)  \right\rangle =2e\frac{d}{dt}\langle n\left(
t\right)  \rangle, \label{Current}%
\end{equation}
where $e$ is the electron charge, $n$ is the number of transferred electrons
across the junction at time $t$, and $\langle n\rangle$ is given by
\begin{equation}
\frac{d}{dt}\left\langle n\left(  t\right)  \right\rangle =i\mathrm{Tr}%
_{osc}\left[  \frac{d}{d\chi}\left(  \frac{d}{dt}\rho_{osc}(\chi,t)\right)
\right]  _{\chi=0}, \label{n}%
\end{equation}
where $\frac{d}{dt}\rho_{osc}(\chi,t)$ is given by Eq. (\ref{master equation}%
). Performing the calculations of Eq.\ (\ref{n}) and simplifying, we obtain
\begin{equation}
\frac{\left\langle I\left(  t\right)  \right\rangle }{2e}=\frac{1}{~2e}\left(
\left\langle I\right\rangle _{0}+\left\langle I\right\rangle _{\mathrm{x}%
}+\left\langle I\right\rangle _{\mathrm{p}}+\left\langle I\right\rangle
_{\mathrm{xp}}+\left\langle I\right\rangle _{\mathrm{q}}\right)  ,
\label{App_current}%
\end{equation}
where physical interpretation of each term is explained in the main text. The
current components are given by%
\begin{equation}
\left\langle I\left(  t\right)  \right\rangle _{0}=2\tilde{\tau}_{0}^{2}%
\xi_{s,\text{ }\sigma}\left(  t\right)  , \label{I0}%
\end{equation}%
\begin{align}
\left\langle I\left(  t\right)  \right\rangle _{\mathrm{x}}  &  =\langle
\hat{x}\rangle\text{ }\frac{2\tilde{\tau}_{0}\tilde{\tau}_{1}}{x_{0}}\left[
\cos\eta\xi_{s,\left(  +\right)  }^{+}\left(  t\right)  -i\sin\eta
\xi_{a,\left(  +\right)  }^{+}\left(  t\right)  \right] \nonumber\\
&  +\langle\hat{x}^{2}\rangle\frac{\tilde{\tau}_{1}^{2}D_{s,\left(  +\right)
}^{+}(t)}{x_{0}^{2}}, \label{Ix}%
\end{align}%
\begin{align}
\text{ }\left\langle I\left(  t\right)  \right\rangle _{\mathrm{p}}  &
=\langle\hat{p}\rangle\frac{2\tilde{\tau}_{0}\tilde{\tau}_{1}}{x_{0}%
M\omega_{0}}~\left[  \cos\eta\xi_{s,\left(  +\right)  }^{-}\left(  t\right)
-i\sin\eta\xi_{a,\left(  +\right)  }^{-}\left(  t\right)  \right] \nonumber\\
&  +\left\langle \hat{p}^{2}\right\rangle \frac{\tilde{\tau}_{1}^{2}%
\gamma_{s,\left(  +\right)  }^{-}\left(  t\right)  }{x_{0}^{2}~M^{~2}%
\omega_{0}^{~2}}, \label{Ip}%
\end{align}%
\begin{equation}
\left\langle I\left(  t\right)  \right\rangle _{\mathrm{xp}}=\left\langle
\left\{  \hat{x},\hat{p}\right\}  \right\rangle \frac{1}{x_{0}^{2}M\omega_{0}%
}\left(  \gamma_{s,\left(  +\right)  }^{+}\left(  t\right)  +D_{s,\left(
+\right)  }^{-}\right)  , \label{Ixp}%
\end{equation}
and%
\begin{equation}
\left\langle I\left(  t\right)  \right\rangle _{\mathrm{q}}=\frac
{2i\hbar\tilde{\tau}_{1}^{2}}{x_{0}^{2}M\omega_{0}}\left(  \gamma_{a,\left(
+\right)  }^{+}\left(  t\right)  -D_{a,\left(  +\right)  }^{-}(t)\right)  .
\label{Iqm}%
\end{equation}
Here, $\left\langle I\right\rangle _{0}$ is the current without the oscillator
coupling, $\left\langle I\right\rangle _{\mathrm{x}}$ and $\left\langle
I\right\rangle _{\mathrm{p}}$ are the currents modulated by the coupled
oscillator through the oscillator's position and momentum coordinates, and
$\ \left\langle I\right\rangle _{\mathrm{q}}\ $is the quantum correction to
the overall current. The terms $\left\langle \hat{x}\right\rangle $,
$\left\langle \hat{p}\right\rangle $ etc. are evaluated by solving Eq.
(\ref{master equation}) using Eq. (\ref{Moments}).

\subsection{Noise\label{AppSec_current_noise}}

The spectral density is
\begin{equation}
\left\langle S\left(  \omega\right)  \right\rangle =2e^{2}\omega\int%
_{0}^{t^{\prime}}dte^{i\omega t}\frac{d}{dt}\left\langle \left\langle
n^{2}\left(  t\right)  \right\rangle \right\rangle ,
\end{equation}
and the covariance $\left\langle \left\langle n^{2}\left(  t\right)
\right\rangle \right\rangle $ is calculated by
\begin{equation}
\frac{d}{dt}\left\langle \left\langle n^{2}\left(  t\right)  \right\rangle
\right\rangle =\frac{d}{dt}\left\langle n^{2}\left(  t\right)  \right\rangle
-2\left\langle n\left(  t\right)  \right\rangle \frac{d}{dt}\left\langle
n\left(  t\right)  \right\rangle .
\end{equation}
All higher moments and the correlations between oscillator coordinates and
transferred charge $n$ are calculated from the master equation
[Eq.\ (\ref{masterAppx})] and by using Eq. (\ref{Moments}). Performing the
calculations and simplifying, we obtain the three noise quantities that depend
on position and momentum of the oscillator,%
\begin{align}
\frac{d}{dt}\left\langle \left\langle n^{2}\left(  t\right)  \right\rangle
\right\rangle  &  =\frac{d}{dt}\left\langle n\left(  t\right)  \right\rangle
+\tilde{N}_{\mathrm{x}}\left(  x,t\right)  +\tilde{N}_{\mathrm{p}}\left(
p,t\right) \nonumber\\
&  +\tilde{N}_{\mathrm{xp}}\left(  x,p,t\right)  , \label{AppCovar}%
\end{align}
where%
\begin{align}
\ddot{N}_{\mathrm{x}}\left(  \hat{x},t\right)   &  =\left\langle \left\langle
\hat{x}n\left(  t\right)  \right\rangle \right\rangle \frac{4\tilde{\tau}%
_{0}\tilde{\tau}_{1}}{x_{0}}\left[  \cos\eta\xi_{s,\left(  +\right)  }%
^{+}\left(  t\right)  -i\sin\eta\xi_{a,\left(  +\right)  }^{+}\left(
t\right)  \right] \nonumber\\
&  +\langle\langle\hat{x}^{2}n\left(  t\right)  \rangle\rangle\frac
{2\tilde{\tau}_{1}^{2}~}{x_{0}^{2}}D_{s,\left(  +\right)  }^{+}(t), \label{Nx}%
\end{align}%
\begin{align}
\ddot{N}_{\mathrm{p}}\left(  \hat{p},t\right)   &  =\left\langle \left\langle
\hat{p}n\left(  t\right)  \right\rangle \right\rangle \frac{4\tilde{\tau}%
_{0}\tilde{\tau}_{1}}{x_{0}M\omega_{0}}\left[  \cos\eta\xi_{s,\left(
+\right)  }^{-}\left(  t\right)  -i\sin\eta\xi_{a,\left(  +\right)  }%
^{-}\left(  t\right)  \right] \nonumber\\
&  +\left\langle \left\langle \hat{p}^{2}n\left(  t\right)  \right\rangle
\right\rangle \frac{2\tilde{\tau}_{1}^{2}~}{x_{0}^{2}M^{2}\omega_{0}^{2}%
}\gamma_{s,\left(  +\right)  }^{-}\left(  t\right)  \label{Np}%
\end{align}%
\begin{equation}
\ddot{N}_{\mathrm{xp}}\left(  \hat{x},\hat{p},t\right)  =\left\langle
\left\langle \left\{  \hat{x},\hat{p}\right\}  n\left(  t\right)
\right\rangle \right\rangle \frac{2\tilde{\tau}_{1}^{2}}{x_{0}^{2}M\omega_{0}%
}\left(  \gamma_{s,\left(  +\right)  }^{+}\left(  t\right)  +D_{s,\left(
+\right)  }^{-}\right)  \label{Nxp}%
\end{equation}
Again, all correlations can be obtained by solving Eq. (\ref{masterAppx}) with
Eq.\ (\ref{Moments}).

Finally, we note the dimensionless scaling used in this paper. The
dimensionless frequency, temperature, applied bias voltage and time are given
by
\begin{equation}
\tilde{\omega}=\frac{\omega}{\omega_{0}},\text{ }\tilde{T}=\frac{k_{B}T}%
{\hbar\omega_{0}},\text{ }\tilde{V}=\frac{eV}{\hbar\omega_{0}},~\tilde
{t}=\omega_{0}t, \label{dW,dT,dV}%
\end{equation}
where $V$ is the applied voltage. The dimensionless position and momentum
operators of the oscillator are given by
\begin{equation}
\tilde{x}\equiv x\left(  \frac{\hbar}{2M\omega_{0}}\right)  ^{-1/2},\text{
}\tilde{p}\equiv p\left(  \frac{\hbar M\omega_{0}}{2}\right)  ^{-1/2}.
\label{dXdp}%
\end{equation}

\section{Details of the terms used in Appx.\ \ref{sec AppendixA}%
\label{AppSec_coeff_list}}%

\begin{align}
A_{1}  &  =\left(  x\rho+\rho x\right)  \frac{\tilde{\tau}_{0}\tilde{\tau}%
_{1}e^{-i\eta}\cos\omega_{0}t^{\prime}}{x_{0}}\\
A_{2}  &  =\left(  p\rho+\rho p\right)  \frac{\tilde{\tau}_{0}\tilde{\tau}%
_{1}e^{-i\eta}\sin\omega_{0}t^{\prime}}{x_{0}M\omega_{0}}%
\end{align}%
\begin{equation}
A_{3}=\left[  \left(  x\rho+\rho x\right)  \cos\omega_{0}t+\left(  p\rho+\rho
p\right)  \frac{\sin\omega_{0}t}{M\omega_{0}}\right]  \frac{\tilde{\tau}%
_{0}\tilde{\tau}_{1}e^{i\eta}}{x_{0}}%
\end{equation}%
\begin{equation}
A_{4}=\left(  px\rho+\rho xp\right)  \frac{\tilde{\tau}_{1}^{2}\sin\omega
_{0}t\cos\omega_{0}t^{\prime}}{x_{0}^{2}M\omega_{0}}%
\end{equation}%
\begin{equation}
A_{5}=\left(  p^{2}\rho+\rho p^{2}\right)  \frac{\tilde{\tau}_{1}^{2}%
\sin\omega_{0}t\sin\omega_{0}t^{\prime}}{x_{0}^{2}M^{2}\omega_{0}^{2}}%
\end{equation}%
\begin{equation}
A_{6}=\left(  x^{2}\rho-2x\rho x+\rho x^{2}\right)  \frac{\tilde{\tau}_{1}%
^{2}\cos\omega_{0}t\cos\omega_{0}t^{\prime}}{x_{0}^{2}}%
\end{equation}%
\begin{equation}
A_{7}=\left(  xp\rho-x\rho p-p\rho x+\rho px\right)  \cos\omega_{0}t\sin
\omega_{0}t^{\prime}%
\end{equation}%
\begin{equation}
A_{8}=2x\rho x\frac{\tilde{\tau}_{1}^{2}\cos\omega_{0}t\cos\omega_{0}%
t^{\prime}}{x_{0}^{2}}%
\end{equation}%
\begin{equation}
A_{9}=\left(  x\rho p+p\rho x\right)  \frac{\tilde{\tau}_{1}^{2}\cos\omega
_{0}t\sin\omega_{0}t^{\prime}}{x_{0}^{2}M\omega_{0}}%
\end{equation}%
\begin{equation}
A_{10}=\rho~\tilde{\tau}_{0}^{2}+x\rho\frac{\tilde{\tau}_{0}\tilde{\tau}%
_{1}e^{-i\eta}\cos\omega_{0}t^{\prime}}{x_{0}}%
\end{equation}%
\begin{equation}
A_{11}=p\rho\frac{\tilde{\tau}_{0}\tilde{\tau}_{1}e^{-i\eta}\sin\omega
_{0}t^{\prime}}{x_{0}M\omega_{0}}%
\end{equation}%
\begin{equation}
A_{12}=\rho x\frac{\tilde{\tau}_{0}\tilde{\tau}_{1}e^{i\eta}\cos\omega_{0}%
t}{x_{0}}+\rho p\frac{\tilde{\tau}_{0}\tilde{\tau}_{1}e^{i\eta}\sin\omega
_{0}t}{x_{0}M\omega_{0}}%
\end{equation}%
\begin{equation}
A_{13}=\left(  x\rho x\cos\omega_{0}t+\frac{x\rho p}{M\omega_{0}}\sin
\omega_{0}t\right)  \frac{\tilde{\tau}_{1}^{2}\cos\omega_{0}t^{\prime}}%
{x_{0}^{2}}%
\end{equation}%
\begin{equation}
A_{14}=\left(  p\rho x\cos\omega_{0}t+\frac{p\rho p}{M\omega_{0}}\sin
\omega_{0}t\right)  \frac{\tilde{\tau}_{1}^{2}\sin\omega_{0}t^{\prime}}%
{x_{0}^{2}M\omega_{0}}%
\end{equation}%
\begin{equation}
A_{15}=\rho x\frac{\tilde{\tau}_{0}\tilde{\tau}_{1}e^{-i\eta}\cos\omega
_{0}t^{\prime}}{x_{0}}%
\end{equation}%
\begin{equation}
A_{16}=\rho p\frac{\tilde{\tau}_{0}\tilde{\tau}_{1}e^{-i\eta}\sin\omega
_{0}t^{\prime}}{x_{0}M\omega_{0}}%
\end{equation}%
\begin{equation}
A_{17}=x\rho~\frac{\tilde{\tau}_{0}\tilde{\tau}_{1}e^{i\eta}\cos\omega_{0}%
t}{x_{0}}%
\end{equation}%
\begin{equation}
A_{18}=p\rho~\frac{\tilde{\tau}_{0}\tilde{\tau}_{1}e^{i\eta}\sin\omega_{0}%
t}{x_{0}M\omega_{0}}%
\end{equation}%
\begin{equation}
A_{19}=\left(  x\rho x\cos\omega_{0}t+\frac{p\rho x}{M\omega_{0}}\sin
\omega_{0}t\right)  \frac{\tilde{\tau}_{1}^{2}\cos\omega_{0}t^{\prime}}%
{x_{0}^{2}}%
\end{equation}%
\begin{equation}
A_{20}=\left(  x\rho p\text{ }\cos\omega_{0}t+\frac{p\rho p}{M\omega_{0}}%
\sin\omega_{0}t\right)  \frac{\tilde{\tau}_{1}^{2}\sin\omega_{0}t^{\prime}%
}{x_{0}^{2}M\omega_{0}}%
\end{equation}

\begin{equation}
B_{1}=\left[  \left(  x\rho-\rho x\right)  \cos\omega_{0}t^{\prime}+\left(
p\rho-\rho p\right)  \frac{\sin\omega_{0}t^{\prime}}{M\omega_{0}}\right]
\frac{\tilde{\tau}_{0}\tilde{\tau}_{1}e^{-i\eta}}{x_{0}}%
\end{equation}%
\begin{equation}
B_{2}=\left[  \left(  x\rho-\rho x\right)  \cos\omega_{0}t+\left(  p\rho-\rho
p\right)  \frac{\sin\omega_{0}t}{M\omega_{0}}\right]  \frac{\tilde{\tau}%
_{0}\tilde{\tau}_{1}e^{i\eta}}{x_{0}}%
\end{equation}%
\begin{equation}
B_{3}=\left(  px\rho-\rho xp\right)  \frac{\tilde{\tau}_{1}^{2}\sin\omega
_{0}t\cos\omega_{0}t^{\prime}}{x_{0}^{2}M\omega_{0}}%
\end{equation}%
\begin{equation}
B_{4}=\left(  p^{2}\rho-\rho p^{2}\right)  \frac{\tilde{\tau}_{1}^{2}%
\sin\omega_{0}t\sin\omega_{0}t^{\prime}}{x_{0}^{2}M^{2}\omega_{0}^{2}}%
\end{equation}%
\begin{equation}
B_{5}=\left(  x^{2}\rho-\rho x^{2}\right)  \frac{\tilde{\tau}_{1}^{2}%
\cos\omega_{0}t\cos\omega_{0}t^{\prime}}{x_{0}^{2}}%
\end{equation}%
\begin{equation}
B_{6}=\left(  p\rho x-x\rho p\right)  \frac{\tilde{\tau}_{1}^{2}\cos\omega
_{0}t\sin\omega_{0}t^{\prime}}{x_{0}^{2}M\omega_{0}}%
\end{equation}%
\begin{equation}
B_{7}=\left(  xp\rho+x\rho p-p\rho x-\rho px\right)  \frac{\tilde{\tau}%
_{1}^{2}\cos\omega_{0}t\sin\omega_{0}t^{\prime}}{x_{0}^{2}M\omega_{0}}%
\end{equation}%
\begin{equation}
B_{8}=\rho\tilde{\tau}_{0}^{2}+\left(  x\rho\cos\omega_{0}t^{\prime}%
+p\rho\frac{\sin\omega_{0}t^{\prime}}{M\omega_{0}}\right)  \frac{\tilde{\tau
}_{0}\tilde{\tau}_{1}e^{-i\eta}}{x_{0}}%
\end{equation}%
\begin{equation}
B_{9}=\left(  \rho x\cos\omega_{0}t+\rho p\frac{\sin\omega_{0}t}{M\omega_{0}%
}\right)  \frac{\tilde{\tau}_{0}\tilde{\tau}_{1}e^{i\eta}}{x_{0}}%
\end{equation}%
\begin{equation}
B_{10}=\left(  x\rho x\cos\omega_{0}t+x\rho p\frac{\sin\omega_{0}t}%
{M\omega_{0}}\right)  \frac{\tilde{\tau}_{1}^{2}\cos\omega_{0}t^{\prime}%
}{x_{0}^{2}}%
\end{equation}

\begin{equation}
B_{11}=\left(  p\rho x\cos\omega_{0}t+p\rho p\frac{\sin\omega_{0}t}%
{M\omega_{0}}\right)  \frac{\tilde{\tau}_{1}^{2}\sin\omega_{0}t^{\prime}%
}{x_{0}^{2}M\omega_{0}}%
\end{equation}%
\begin{equation}
B_{12}=\left(  \rho x\cos\omega_{0}t^{\prime}+\rho p\frac{\sin\omega
_{0}t^{\prime}}{M\omega_{0}}\right)  \frac{\tilde{\tau}_{0}\tilde{\tau}%
_{1}e^{-i\eta}}{x_{0}}%
\end{equation}%
\begin{equation}
B_{13}=\left(  x\rho\cos\omega_{0}t+p\rho\frac{\sin\omega_{0}t}{M\omega_{0}%
}\right)  \frac{\tilde{\tau}_{0}\tilde{\tau}_{1}e^{i\eta}}{x_{0}}%
\end{equation}%
\begin{equation}
B_{14}=\left(  x\rho x\cos\omega_{0}t+p\rho x\frac{\sin\omega_{0}t}%
{M\omega_{0}}\right)  \frac{\tilde{\tau}_{1}^{2}\cos\omega_{0}t^{\prime}%
}{x_{0}^{2}}%
\end{equation}%
\begin{equation}
B_{15}=\left(  x\rho p\cos\omega_{0}t+p\rho p\frac{\sin\omega_{0}t}%
{M\omega_{0}}\right)  \frac{\tilde{\tau}_{1}^{2}\sin\omega_{0}t^{\prime}%
}{x_{0}^{2}M\omega_{0}}%
\end{equation}

\begin{equation}
C_{1}=\left(  x\rho+\rho x\right)  \frac{\tilde{\tau}_{0}\tilde{\tau}%
_{1}e^{i\eta}\cos\omega_{0}t^{\prime}}{x_{0}}%
\end{equation}%
\begin{equation}
C_{2}=\left(  p\rho+\rho p\right)  \frac{\tilde{\tau}_{0}\tilde{\tau}%
_{1}e^{i\eta}\sin\omega_{0}t^{\prime}}{x_{0}M\omega_{0}}%
\end{equation}%
\begin{equation}
C_{3}=\left[  \left(  x\rho+\rho x\right)  \cos\omega_{0}t+\left(  p\rho+\rho
p\right)  \frac{\sin\omega_{0}t}{M\omega_{0}}\right]  \frac{\tilde{\tau}%
_{0}\tilde{\tau}_{1}e^{-i\eta}}{x_{0}}%
\end{equation}%
\begin{equation}
C_{4}=\left(  px\rho+\rho xp\right)  \frac{\tilde{\tau}_{1}^{2}\sin\omega
_{0}t\cos\omega_{0}t^{\prime}~}{x_{0}^{2}M\omega_{0}}%
\end{equation}%
\begin{equation}
C_{5}=\left(  p^{2}\rho+\rho p^{2}\right)  \frac{\tilde{\tau}_{1}^{2}%
\sin\omega_{0}t\sin\omega_{0}t^{\prime}}{x_{0}^{2}M^{2}\omega_{0}^{2}}%
\end{equation}%
\begin{equation}
C_{6}=\left(  x^{2}\rho-2x\rho x+\rho x^{2}\right)  \frac{\tilde{\tau}_{1}%
^{2}\cos\omega_{0}t\cos\omega_{0}t^{\prime}}{x_{0}^{2}}%
\end{equation}%
\begin{equation}
C_{7}=\left(  xp\rho-x\rho p-p\rho x+\rho px\right)  \frac{\tilde{\tau}%
_{1}^{2}\cos\omega_{0}t\sin\omega_{0}t^{\prime}}{x_{0}^{2}M\omega_{0}}%
\end{equation}%
\begin{equation}
C_{8}=2x\rho x\frac{\tilde{\tau}_{1}^{2}\cos\omega_{0}t\cos\omega_{0}%
t^{\prime}}{x_{0}^{2}}%
\end{equation}%
\begin{equation}
C_{9}=\left(  x\rho p+p\rho x\right)  \frac{\tilde{\tau}_{1}^{2}\cos\omega
_{0}t\sin\omega_{0}t^{\prime}}{x_{0}^{2}M\omega_{0}}%
\end{equation}%
\begin{equation}
C_{10}=\left(  \rho x\cos\omega_{0}t^{\prime}+\rho p\frac{\sin\omega
_{0}t^{\prime}}{M\omega_{0}}\right)  \frac{\tilde{\tau}_{0}\tilde{\tau}%
_{1}e^{i\eta}}{x_{0}}%
\end{equation}%
\begin{equation}
C_{11}=\left(  x\rho\cos\omega_{0}t+p\rho\frac{\sin\omega_{0}t}{M\omega_{0}%
}\right)  \frac{\tilde{\tau}_{0}\tilde{\tau}_{1}e^{-i\eta}}{x_{0}}%
\end{equation}%
\begin{equation}
C_{12}=\left(  x\rho x\cos\omega_{0}t+p\rho x\frac{\sin\omega_{0}t}%
{M\omega_{0}}\right)  \frac{\tilde{\tau}_{1}^{2}\cos\omega_{0}t^{\prime}%
}{x_{0}^{2}}%
\end{equation}%
\begin{equation}
C_{13}=\left(  x\rho p\cos\omega_{0}t+p\rho p\frac{\sin\omega_{0}t}%
{M\omega_{0}}\right)  \frac{\tilde{\tau}_{1}^{2}\sin\omega_{0}t^{\prime}%
}{x_{0}^{2}M\omega_{0}}%
\end{equation}%
\begin{equation}
C_{14}=\left(  x\rho\cos\omega_{0}t^{\prime}+p\rho\frac{\sin\omega
_{0}t^{\prime}}{M\omega_{0}}\right)  \frac{\tilde{\tau}_{0}\tilde{\tau}%
_{1}e^{i\eta}}{x_{0}}%
\end{equation}%
\begin{equation}
C_{15}=\left(  \rho x\cos\omega_{0}t+\rho p\frac{\sin\omega_{0}t}{M\omega_{0}%
}\right)  \frac{\tilde{\tau}_{0}\tilde{\tau}_{1}e^{-i\eta}}{x_{0}}%
\end{equation}%
\begin{equation}
C_{16}=\left(  x\rho x\cos\omega_{0}t+x\rho p\frac{\sin\omega_{0}t}%
{M\omega_{0}}\right)  \frac{\tilde{\tau}_{1}^{2}\cos\omega_{0}t^{\prime}%
}{x_{0}^{2}}%
\end{equation}%
\begin{equation}
C_{17}=\left(  p\rho x\cos\omega_{0}t+p\rho p\frac{\sin\omega_{0}t}%
{M\omega_{0}}\right)  \frac{\tilde{\tau}_{1}^{2}\sin\omega_{0}t^{\prime}%
}{x_{0}^{2}M\omega_{0}}%
\end{equation}%
\begin{equation}
D_{1}=\left[  \left(  x\rho-\rho x\right)  \frac{\tilde{\tau}_{0}\tilde{\tau
}_{1}e^{i\eta}\cos\omega_{0}t^{\prime}}{x_{0}}\right]
\end{equation}%
\begin{equation}
D_{2}=\left(  p\rho-\rho p\right)  \frac{\tilde{\tau}_{0}\tilde{\tau}%
_{1}e^{i\eta}\sin\omega_{0}t^{\prime}}{x_{0}M\omega_{0}}%
\end{equation}%
\begin{equation}
D_{3}=\left[  \left(  x\rho-\rho x\right)  \cos\omega_{0}t+\left(  p\rho-\rho
p\right)  \frac{\sin\omega_{0}t}{M\omega_{0}}\right]  \frac{\tilde{\tau}%
_{0}\tilde{\tau}_{1}e^{-i\eta}}{x_{0}}%
\end{equation}%
\begin{equation}
D_{4}=\left(  px\rho-\rho xp\right)  ~\frac{\tilde{\tau}_{1}^{2}\sin\omega
_{0}t\cos\omega_{0}t^{\prime}}{x_{0}^{2}M\omega_{0}}%
\end{equation}%
\begin{equation}
D_{5}=\left(  p^{2}\rho-\rho p^{2}\right)  ~\frac{\tilde{\tau}_{1}^{2}%
\sin\omega_{0}t\sin\omega_{0}t^{\prime}}{x_{0}^{2}M^{2}\omega_{0}^{2}}%
\end{equation}%
\begin{equation}
D_{6}=\left(  x^{2}\rho-\rho x^{2}\right)  \text{ }\frac{\tilde{\tau}_{1}%
^{2}\cos\omega_{0}t\cos\omega_{0}t^{\prime}}{x_{0}^{2}}%
\end{equation}%
\begin{equation}
D_{7}=\left(  p\rho x-x\rho p\right)  ~\frac{\tilde{\tau}_{1}^{2}\cos
\omega_{0}t\sin\omega_{0}t^{\prime}}{x_{0}^{2}M\omega_{0}}%
\end{equation}%
\begin{equation}
D_{8}=\left(  xp\rho+x\rho p-p\rho x-\rho px\right)  \frac{\tilde{\tau}%
_{1}^{2}\cos\omega_{0}t\sin\omega_{0}t^{\prime}}{x_{0}^{2}M\omega_{0}}%
\end{equation}%
\begin{equation}
D_{9}=\left(  \rho x\cos\omega_{0}t^{\prime}+\rho p\frac{\sin\omega
_{0}t^{\prime}}{M\omega_{0}}\right)  \frac{\tilde{\tau}_{0}\tilde{\tau}%
_{1}e^{i\eta}}{x_{0}}%
\end{equation}%
\begin{equation}
D_{10}=\left(  x\rho\cos\omega_{0}t+p\rho\frac{\sin\omega_{0}t}{M\omega_{0}%
}\right)  \frac{\tilde{\tau}_{0}\tilde{\tau}_{1}e^{-i\eta}}{x_{0}}%
\end{equation}%
\begin{equation}
D_{11}=\left(  x\rho x\cos\omega_{0}t+p\rho x\frac{\sin\omega_{0}t}%
{M\omega_{0}}\right)  \frac{\tilde{\tau}_{1}^{2}\cos\omega_{0}t^{\prime}%
}{x_{0}^{2}}%
\end{equation}%
\begin{equation}
D_{12}=\left(  x\rho p\cos\omega_{0}t+p\rho p\frac{\sin\omega_{0}t}%
{M\omega_{0}}\right)  \frac{\tilde{\tau}_{1}^{2}\sin\omega_{0}t^{\prime}%
}{x_{0}^{2}M\omega_{0}}%
\end{equation}%
\begin{equation}
D_{13}=\left(  x\rho\cos\omega_{0}t^{\prime}+p\rho\frac{\sin\omega
_{0}t^{\prime}}{M\omega_{0}}\right)  \frac{\tilde{\tau}_{0}\tilde{\tau}%
_{1}e^{i\eta}}{x_{0}}%
\end{equation}%
\begin{equation}
D_{14}=\left(  \rho x\cos\omega_{0}t+\rho p\frac{\sin\omega_{0}t}{M\omega_{0}%
}\right)  \frac{\tilde{\tau}_{0}\tilde{\tau}_{1}e^{-i\eta}}{x_{0}}%
\end{equation}%
\begin{equation}
D_{15}=\left(  x\rho x\cos\omega_{0}t+p\rho x\frac{\sin\omega_{0}t}%
{M\omega_{0}}\right)  \frac{\tilde{\tau}_{1}^{2}\cos\omega_{0}t^{\prime}%
}{x_{0}^{2}}%
\end{equation}%
\begin{equation}
D_{16}=\left(  x\rho p\cos\omega_{0}t+p\rho p\frac{\sin\omega_{0}t}%
{M\omega_{0}}\right)  \frac{\tilde{\tau}_{1}^{2}\sin\omega_{0}t^{\prime}%
}{x_{0}^{2}M\omega_{0}}%
\end{equation}


\begin{thebibliography}{99}                                                                                               %


\bibitem {CJG11}P.-W. Chen, C.-C. Jian and H.-S. Goan , Phys. Rev. B
\textbf{83}, 115439 (2011).

\bibitem {STB10}J. Stettenheim, M. Thalakulam, F. Pan, M. Bal, Z. Ji, W. Xue,
L. Pfeiffer, K.W. West, M.P. Blencowe and A.J. Rimberg, Nature \textbf{466},
09123 p86--90 (2010).

\bibitem {OYN09}S. H. Ouyang, J. Q. You and F. Nori, Phys. Rev. B \textbf{79},
075304 (2009).

\bibitem {SWN07}L. F. Wei, Yu-xi Liu, C. P. Sun, and Franco Nori Phys. Rev.
Lett. \textbf{97}, 237201 (2006).

\bibitem {UGM04}D. W. Utami, H.-S. Goan, and G. J. Milburn, Phys. Rev. B
\textbf{70}, 075303 (2004).

\bibitem {CB04}N. M. Chtchelkatchev, W. Belzig and C. Bruder, Phys. Rev. B
\textbf{70}, 193305 (2004).

\bibitem {ZB02}Y. Zhang and M. P. Blencowe, J. Appl. Phys. \textbf{91}, 4249 (2002).

\bibitem {EB01}A. Erbe, C. Weiss, W. Zwerger and R. H. Blick, Phys. Rev. Lett.
\textbf{87}, 096106 (2001).

\bibitem {Graig00}H. G. Graighead, Science \textbf{290}, 1532 (2000).

\bibitem {SLN13}D. H. Santamore, Neill Lambert, Franco Nori, Phys. Rev. B 87,
075422 (2013).

\bibitem {Arm04}A. D. Armour, Phys. Rev. B \textbf{70}, 165315 (2004).

\bibitem {KBV06}F. H. L. Koppens, C. Buizert, K. J. Tielrooij, I.T. Vink, K.C.
Nowack, T. Meunier, L. P. Kouwenhoven and L. M. K. Vandersypen, Nature
\textbf{442}, 766 (2006).

\bibitem {AS05}D. V. Averin and E. V. Sukhorukov, Phys. Rev. Lett.
\textbf{95},126803 (2005).

\bibitem {PH05}M. Pioro-Ladriere, R. Abolfath, P. Zawadzki, J. Lapointe, S. A.
Studenikin, A. S. Sachrajda and P. Hawrylak, Phys. Rev. B \textbf{72}, 125307 (2005).

\bibitem {PG05}J. R. Petta, A. C. Johnson, J. M. Taylor, E. A. Laird, A.
Yacoby, M. D. Lukin, C. M. Marcus, M. P. Hanson and A. C. Gossard, Science
\textbf{309}, 2180 (2005).

\bibitem {PB02}S. Pilgram and M. Buttiker, Phys. Rev. Lett. \textbf{89},
200401 (2002).

\bibitem {WKR05}J. Wabnig, D. V. Khomitsky, J. Rammer and A. L. Shelankov,
Phys. Rev. B. \textbf{72} 165347 (2005).

\bibitem {GC05}M. R. Geller and A. N. Cleland, Phys. Rev. A \textbf{71},
032311 (2005).

\bibitem {MSS01}Y. Makhlin, G. Schon and A. Shnirman, Rev. Mod. Phys.
\textbf{73}, 357 (2001).

\bibitem {PJR08}M. Poggio, M. P. Jura, C. L. Degen, M. A. Topinka, H. J.
Mamin, D. Goldhaber- Gordon and D. Rugar, Nature Phys. \textbf{4}, 635 (2008).

\bibitem {BB12}L. L. Benatov and M. P. Blencowe, Phys. Rev. B \textbf{86},
075313 (2012).

\bibitem {CG04}A. A. Clerk and S. M. Girvin, Phys. Rev. B \textbf{70}, 121303 (2004).

\bibitem {Clerk04b}A. Clerk, Phys. Rev. B \textbf{70}, 245306 (2004).

\bibitem {Rammer04}J. Rammer, A. L. Shelankov and J. Wabnig Phys. Rev. B
\textbf{70},115327 (2004).

\bibitem {BC08}S. D. Bennett and A. A. Clerk, Phys. Rev. B \textbf{78}, 165328 (2008).

\bibitem {DTB07}C. B. Doiron, B. Trauzettel and C. Bruder, Phys. Rev. B
\textbf{76}, 195312 (2007).

\bibitem {Breuer10}H. P. Breuer and F. Petruccione, \textit{The Theory of Open
Quantum Systems}, (Oxford University Press, Oxford U.K, 2002)

\bibitem {M49}D. K. C. MacDonald, Rep. Prog. Phys.,\textbf{12}, 56 (1949).

\bibitem {WT11}S. Walter and B. Trauzettel, Phys. Rev. B \textbf{83}, 155411 (2011).

\bibitem {TM74}H. Takahasi and M. Mori, Research Institute for Mathematical
Science (RIMS), \textbf{9}, 721, (1974)

\bibitem {OM99}T. Ooura and M. Mori., J. Comput. Appl. Math. 112, 229 (1999).

\bibitem {SMH03}Anatoly Yu. Smirnov, Lev G. Mourokh, and Norman J. M. Horing,
Phys. Rev. B \textbf{67}, 115312 (2003).
\end{thebibliography}
\end{document}